\documentclass[aps,pre,twocolumn,groupedaddress]{revtex4-1}
\usepackage{graphicx}
\usepackage{amsmath}
\usepackage{amssymb}

\begin{document}

\title{Full-order fluctuation-dissipation relation for a class of nonequilibrium steady states}

\author{Akihisa Ichiki}
\email{ichiki@gvm.nagoya-u.ac.jp}
\affiliation{Green Mobility Collaborative Research Center, Nagoya University, Furo-cho, Chikusa-ku, Nagoya 464-8603, Japan}
\author{Masayuki Ohzeki}
\email{mohzeki@i.kyoto-u.ac.jp}
\affiliation{Department of Systems Science, Graduate School of Information, Kyoto University, Yoshida-Honmachi, Sakyo-ku, Kyoto 606-8501, Japan}
\date{\today}

\begin{abstract}
Acceleration of relaxation toward a fixed stationary distribution via violation of detailed balance was reported in the context of a Markov chain Monte Carlo method recently.  
Inspired by this result, systematic methods to violate detailed balance in Langevin dynamics were formulated by using exponential and rotational nonconservative forces.  
In the present paper, we accentuate that such specific nonconservative forces relate to the large deviation of total heat in an equilibrium state.  
The response to these nonconservative forces can be described by the intrinsic fluctuation of the total heat in the equilibrium state.  
Consequently, the fluctuation-dissipation relation for nonequilibrium steady states is derived without recourse to a linear response approximation.  
\end{abstract}

\maketitle

\section{Introduction}
One of the focal topics in nonequilibrium thermodynamics and statistical physics is an extension of the fluctuation-dissipation theorem (FDT) \cite{Kubo}.  
The well-known FDT for equilibrium states owes its success to two relations: a {\it fluctuation-response relation} and a {\it relation between the response and energy dissipation}.  
The fluctuation-response relation claims that the response of a macroscopic quantity to a perturbation is given by its intrinsic fluctuation under the unperturbed dynamics.  
On the other hand, for near-equilibrium state, a linear response to a small perturbation can be expected.  The energy dissipation is hence given in terms of the response.  
In far-from-equilibrium states, however, a general relation between energy dissipation and the response is missing, while several extensions of the fluctuation-response relation are known \cite{Prost, Seifert, Baiesi}.  
Only in the linear response regime near nonequilibrium steady states (NESS), where the energy dissipation can be macroscopically evaluated, has the genuine fluctuation-dissipation relation been obtained \cite{Harada, Harada2}.  

Highlighting a series of Onsager's works \cite{Onsager, Onsager2}, Onsager's regression hypothesis gives the fluctuation-response relation in near-equilibrium states.  
Onsager's regression hypothesis assumes that the correlation function in equilibrium state ({\it intrinsic fluctuation}) governs the relaxation from near-equilibrium state toward equilibrium that is given as the {\it response} to an appropriate impulse perturbation. 
On the other hand, Onsager's principle provides a relationship between {\it response} to an external perturbation sustaining a steady state and {\it energy dissipation} in the near-equilibrium steady state, which is housekeeping heat characterizing violation of the detailed balance condition (DBC).
Onsager's principle states that the current as a response is proportional to the external perturbation called thermodynamic force. 
Thus the housekeeping heat that is given by a product of the current and its conjugate thermodynamic force is always non-negative.
Combining Onsager's regression hypothesis (fluctuation-response relation) and Onsager's principle (relation between response and dissipation) yields FDT in near-equilibrium state. 
Recently, the influence of the violation of the DBC on relaxation has been discussed in the context of a Markov chain Monte Carlo method (MCMC) \cite{IchikiOhzeki}. 
It is guaranteed in terms of eigenvalues for a transition matrix that the relaxation toward a fixed stationary distribution is accelerated by the violation of the DBC \cite{IchikiOhzeki}.  
Furthermore, it has been found that applying exponential and/or rotational forces as a systematic method to violate the DBC accelerates relaxation in Langevin dynamics \cite{OhzekiIchiki,OhzekiIchiki2}.  
In these works on MCMC, the energy dissipation is obviously given by a housekeeping heat, which is a product of probability current and nonconservative force violating DBC. 
If FDT in NESS far from equilibrium is expected, relaxation (response) should be connected with intrinsic fluctuation. 
However, what is the intrinsic fluctuation relating the relaxation or response in this case? 
In the present paper, we give a possible answer to this question. 

The main result of this paper is to give the relation between the large deviation function for total heat in an equilibrium state and the expectation of that in the NESS.  
The large deviation function describes the intrinsic fluctuation in the equilibrium state.  
In addition, the housekeeping heat, which is equivalent to the total heat in the NESS, expresses the energy dissipation.  
Therefore, our result is interpreted as the extension of the FDT in the NESS.  
The result is derived without resorting to a linear response approximation as used in several extensions of the FDT in the NESS.  
In other words, our extension leads to a full-order form of the FDT in the NESS.  
The fluctuation-response relation for the total heat is derived through the framework of the Nemoto-Sasa theory \cite{NemotoSasa, SughiyamaOhzeki, SughiyamaOhzeki2011}, which gives the relation between the large deviation function for intrinsic fluctuations and response to appropriate perturbations.  
The response part of this relation 
contains the housekeeping heat emerging from a topological argument, which is the general framework given by Sagawa and Hayakawa \cite{Sagawa}.  

\section{Set up} 
In the present paper, we deal with the two systems, which are related by the additional force ${\bf u}$ on the $N$ degrees of freedom. 
We refer the case without any additional force as the original system and that with nontrivial ${\bf u}$ as the biased system.
The system is governed by the following Langevin dynamics 
\begin{eqnarray}
d{\bf x}(t)=\left[{\bf A}\left(\bar{\bf x}(t)\right)+{\bf u}\left(\bar{\bf x}(t)\right)\right] dt + \sqrt{2T}d{\bf W}(t),\label{u_Langevin}
\end{eqnarray}
where ${\bf A}$ is the drift term for the original system, ${\bf W}(t)$ is the standard Wienner process, $T$ denotes the noise intensity or temperature, and a midpoint prescription $\bar{\bf x}(t)=\left[{\bf x}(t+dt)+{\bf x}(t)\right]/2$ is used for Stratonovich interpretation of stochastic dynamics.  

The housekeeping heat for the dynamics (\ref{u_Langevin}) is defined as \cite{HatanoSasa}
\begin{eqnarray}
Q_{\rm hk}=\int_0^\tau \left({\bf A}+{\bf u}-T{\rm grad} \ln P_{\rm ss}^{\bf u}\right)\circ\dot{\bf x}(t) dt,
\end{eqnarray}
where $P_{\rm ss}^{\bf u}$ is the stationary distribution and $\circ$ stands for the multiplication in the sense of the Stratonovich.  
On the other hand, in NESS, the excess heat is defined as 
\begin{eqnarray}
Q_{\rm ex}= -T\ln P_{\rm ss}^{\bf u}\left( {\bf x}(\tau)\right)+T\ln P_{\rm ss}^{\bf u}\left({\bf x}(0)\right).
\end{eqnarray}
Using these quantities, the total heat is defined as $Q_{\rm tot}=Q_{\rm hk}+Q_{\rm ex}$.  
Since $\left({\bf A}+{\bf u}-T{\rm grad} \ln P_{\rm ss}^{\bf u}\right)P_{\rm ss}^{\bf u}$ is the probability current, which characterizes the violation of DBC, the housekeeping heat $Q_{\rm hk}$ vanishes if the DBC is satisfied. 

\section{Nemoto-Sasa theory and variational principle}
For convenience, we briefly review the formulation of the Nemoto-Sasa theory from the viewpoint of a variational principle \cite{SughiyamaOhzeki, SughiyamaOhzeki2011}.  
The Nemoto-Sasa theory originally gives the relation between the cumulant generating function of a current in the NESS of the original system and an expectation of the current in the biased system \cite{NemotoSasa}. 
The conditional path probability for a path realization ${\bf X}$ with an initial condition ${\bf x}(0)={\bf x}_0$ is given as 
\begin{eqnarray}
&&\hspace{-8mm}L_{\bf u}\left({\bf X}|{\bf x}_0\right)\nonumber\\
&\propto& \exp\Big\{-\frac{1}{4T}\int_0^\tau dt\left[\dot{\bf x}(t)-{\bf A}\left(\bar{\bf x}(t)\right)-{\bf u}\left(\bar{\bf x}(t)\right)\right]^2\nonumber\\
&&-\frac{1}{2}\int_0^\tau dt\, {\rm div}\left[{\bf A}\left(\bar{\bf x}(t)\right)+{\bf u}\left(\bar{\bf x}(t)\right)\right]\Big\}.\label{path_probability}
\end{eqnarray}
Note that 
\begin{eqnarray}
\ln\frac{L_{\bf u}({\bf X}|{\bf x}_0)}{L_{\bf 0}({\bf X}|{\bf x}_0)}&=&\frac{1}{2T}\int_0^\tau dt\left[ {\bf u}\cdot({\bf A}+{\bf u})-(2{\bf A}+{\bf u})\cdot\frac{\bf u}{2}\right]\nonumber\\
&-&\frac{1}{2}\int_0^\tau dt\, {\rm div}{\bf u}+\int_0^\tau \frac{\bf u}{\sqrt{2T}}\circ d{\bf W}\nonumber\\
&&\hspace{-17mm}=\int_0^\tau\frac{{\bf u}^2}{4T}dt -\int_0^\tau \frac{{\rm div}{\bf u}}{2}dt + \int_0^\tau \frac{\bf u}{\sqrt{2T}}\circ d{\bf W}(t).\label{suppl}
\end{eqnarray} 
Since the expectation of the second term in the last line of Eq.~(\ref{suppl}) is canceled by that of the third term, we find the scaled Kullback-Leibler (KL) divergence between path probabilities with and without the additional force ${\bf u}$ as 
\begin{eqnarray}
D[L_{\bf u}|L_{\bf 0}]&\equiv&\displaystyle\lim_{\tau\to\infty}\left\langle\frac{1}{\tau}\ln\frac{L_{\bf u}({\bf X}|{\bf x}_0)}{L_{\bf 0}({\bf X}|{\bf x}_0)}\right\rangle_{\bf u}\nonumber\\
&=&\int d{\bf x} \frac{{\bf u}^2({\bf x})}{4T}P_{\rm ss}^{\bf u}({\bf x}),\label{KL}
\end{eqnarray}
where $\langle\cdot\rangle_{\bf u}$ denotes the ensemble average under the stochastic dynamics (\ref{u_Langevin}).   To investigate the full-order cumulant, we recall the scaled cumulant generating function $\lambda_0(\gamma)$ for an arbitrary time-averaged quantity $S({\bf X})$ of the original system defined as 
\begin{eqnarray}
\lambda_0(\gamma)\equiv\displaystyle\lim_{\tau\to\infty}\frac{1}{\tau}\ln\left\langle\exp\left(\gamma\tau S({\bf X})\right)\right\rangle_{\bf 0}.
\end{eqnarray}
Here let us minimize $D[L_{\bf u}|L_{\bf 0}]$ under the constraint that the expectation $\langle S({\bf X})\rangle_{\bf u}$ depending on the path realization ${\bf X}$ is fixed.
The scaled cumulant generating function then emerges as
\begin{eqnarray}
\lambda_0(\gamma) = \displaystyle\max_{\bf u}\left\{\gamma\left\langle S({\bf X})\right\rangle_{\bf u}-\int d{\bf x}\frac{{\bf u}^2({\bf x})}{4T}P_{\rm ss}^{\bf u}({\bf x})\right\},\label{lam}
\end{eqnarray}
where we set $\gamma$ as a Lagrange multiplier \cite{SughiyamaOhzeki}. 
The cumulant generating function satisfies the following fluctuation-response relation: 
\begin{eqnarray}
\frac{\partial\lambda_0(\gamma)}{\partial \gamma}=\left\langle S({\bf X})\right\rangle_{{\bf u}^\gamma},\label{FRR}
\end{eqnarray}
where the special additional force ${\bf u}^\gamma$ is given by
\begin{eqnarray}
{\bf u}^\gamma = \arg\displaystyle\max_{\bf u}\left\{\gamma\left\langle S({\bf X})\right\rangle_{\bf u}-\int d{\bf x}\frac{{\bf u}^2({\bf x})}{4T}P_{\rm ss}^{\bf u}({\bf x})\right\}.\label{u}
\end{eqnarray}
The fluctuation-response relation indicates that the cumulant generating function in the original system can be estimated through the measurement of the quantity $S({\bf X})$ in the biased system.
 
On the other hand, the large deviation function of $S({\bf X})$ of the original system \begin{eqnarray}
I_0(s)\equiv -\displaystyle\lim_{\tau\to\infty}\frac{1}{\tau}\ln\left[{\rm Prob}_{\bf 0}\left(S({\bf X}) = s\right)\right],
\end{eqnarray}
where ${\rm Prob}_{\bf 0}\left(S({\bf X}) = s\right)$ denotes the probability that $S({\bf X}) = s$ in the original system, is given by the minimum of the KL divergence as 
\begin{eqnarray}
I_0\left(s\right) = \displaystyle\min_{\bf u} D\left[L_{\bf u}|L_{\bf 0}\right] \quad{\rm subject}\,\,{\rm to }\quad S\left({\bf X}\right)=s.\label{large_deviation}
\end{eqnarray}
This relation is immediately obtained by the Legendre transformation on Eq. (\ref{lam}).
The special additional force gives the solution of this equality.
In other words, the large deviation function of the original system can be evaluated through the biased system as shown in Eq. (\ref{KL}).
Substitution of $S({\bf X})=\int_0^\tau \dot{x}dt/\tau$ into Eqs.~(\ref{lam}) and (\ref{u}) indeed reproduces the Nemoto-Sasa theory for the current cumulant generating function \cite{NemotoSasa}.  
  Note that the above formulation is valid for an arbitrary quantity $S({\bf X})$, not only for the current.  

\section{Two choices of DBC violation in MCMC}
The MCMC is a powerful tool for providing a sequence of random numbers following a desired distribution.
Various techniques to make the relaxation to the stationary state faster have been proposed \cite{Swendsen, HukushimaNemoto, Neal, Ohzeki,Ohzeki2}.  
While these conventional MCMCs impose the DBC to ensure the convergence of the system toward the desired distribution, the DBC is only the sufficient condition for the convergence.  
Several algorithms violating the DBC actually exhibit high-speed convergences toward the desired distribution \cite{SuwaTodo, Turitsyn, Fernandes, SakaiHukushima}.  
General proof for this acceleration has been mathematically provided \cite{IchikiOhzeki}.  
Furthermore, a systematic method to violate the DBC \cite{OhzekiIchiki,OhzekiIchiki2} is provided as follows.

The desired distribution $P_{{\rm ss}}$ in the MCMC often takes the form of an exponential family, or in physical terminology, a Gibbsian distribution.  
It is wellknown that such a distribution is realized as an equilibrium distribution by the following Langevin dynamics with the DBC: 
\begin{eqnarray}
d{\bf x}(t) = -{\rm grad} U\left(\bar{{\bf x}}(t)\right) dt + \sqrt{2T}d{\bf W}(t).\label{DBC_Langevin}
\end{eqnarray}
Here $U$ is a scalar potential given as the summation of the ingredients associated with the $i$th degree of freedom $U_i({\bf x})$ as $U({\bf x})=\sum_i U_i({\bf x})$, and $T=1/\beta$ temperature.  
The equilibrium distribution for this dynamics is given as 
\begin{eqnarray}
P_{\rm ss}({\bf x})=\exp\left[-\beta U\left({\bf x}\right)\right]/Z,\label{Gibbs}
\end{eqnarray}
where $Z$ is a partition function.  
In the systematic method violating DBC while keeping the Gibbsian distribution (\ref{Gibbs}) as a stationary one, we add a non-conservative force ${\bf u}$ to the dynamics (\ref{DBC_Langevin}): 
\begin{eqnarray}
d{\bf x}(t) = -{\rm grad} U\left(\bar{{\bf x}}(t)\right) dt + {\bf u}\left(\bar{\bf x}(t)\right) dt + \sqrt{2T}d{\bf W}(t).\nonumber\\\label{DBC_violating_Langevin}
\end{eqnarray}
According to the Fokker-Planck equation corresponding to Eq.~(\ref{DBC_violating_Langevin}), we heuristically find two solutions for ${\bf u}$ to keep the Gibbsian distribution (\ref{Gibbs}) as a stationary distribution for the modified dynamics \cite{OhzekiIchiki,OhzekiIchiki2}: 
(i) Rotational force is given by 
\begin{eqnarray}\nonumber
u_i\left({\bf x}\right)=\displaystyle\left[\sum_{j(<i)}\gamma_{ij}({\bf x}_{\backslash i,j})\frac{\partial U({\bf x})}{\partial x_j}-\sum_{j(>i)}\gamma_{ij}({\bf x}_{\backslash i,j})\frac{\partial U({\bf x})}{\partial x_j}\right],
\\
\label{rotational}
\end{eqnarray}
where $x_i$ and $u_i$ are the $i$th components of ${\bf x}$ and ${\bf u}$, respectively, and $\gamma_{ij}({\bf x}_{\backslash i,j})$ is an arbitrary antisymmetric matrix independent of $x_i$ and $x_j$.
Here ${\bf x}_{\backslash i, j}$ denotes an $(N-2)$-dimensional subvector given by the elimination of the $i$th and $j$th components from ${\bf x}$.
(ii) Exponential force is 
\begin{eqnarray}
u_i\left({\bf x}\right)=\gamma_i({\bf x}_{\backslash i}) \exp\left[\beta U_i\left(x_i\right)\right],\label{exponential}
\end{eqnarray}
where $\gamma_i ({\bf x}_{\backslash i})$ is an arbitrary function independent of the $i$th component $x_i$.
The independence of $x_{i}$ and $x_j$ in the constant $\gamma_{ij}$ and that of $x_{i}$ in $\gamma_i$ in the additional force comes from the condition that fixes the stationary distribution.
For the exponential force (\ref{exponential}), a periodic boundary condition should be imposed due to the probability conservation, while the original dynamics (\ref{DBC_Langevin}) can be solved under the natural boundary condition.  

\section{Rotational force}
The additional force ${\bf u}$ violating the DBC was first formulated to accelerate the relaxation to the stationary distribution for practical use in numerical computation.
On the other hand, as shown below, it naturally emerges from the optimization of the KL divergence.  
Let us consider the long-time average of work performed by the external force ${\bf A}+{\bf u}$ with ${\bf A}=-{\rm grad} U$: 
\begin{eqnarray}
\overline{{W}}_{\bf u}=\displaystyle\lim_{\tau\to\infty}\frac{1}{\tau}\int_C \left[{\bf A}({\bf x})+{\bf u}({\bf x})\right]\cdot d\vec{\ell},
\end{eqnarray}
where the integral is performed on the trajectory $C$ in the state space, which is given by the path realization ${\bf X}$.
Supposing that the trajectory $C$ is compact and no source exists, $\overline{{W}}_{\bf u}$ vanishes for any open trajectories $C$.
This assumption is valid if no divergence of the probability current exists in the stationary state.
Therefore we focus on the case only for closed trajectories.
Using the Stokes theorem, we find 
\begin{eqnarray}
\overline{W}_{\bf u}=\displaystyle\lim_{\tau\to\infty}\sum_{i, j}\frac{N_{ij}}{\tau}\int_D \frac{\partial \left[A_i + u_i\right]}{\partial x_j}dx_i \wedge dx_j,
\end{eqnarray}
where the integration is carried out in the region $D$ whose boundary is $C$, i.e., $C=\partial D$, $N_{ij}$ is the number of rotations in the $i$-$j$ plane, and $\wedge$ denotes a wedge product.
Note that $N_{ij}$ depends on the location, but is independent of $x_i$ and $x_j$.  
Then the time derivative of housekeeping heat, which coincides with $\left\langle \overline{{W}}_{\bf u}\right\rangle_{\bf u}$ because of stationarity of the system after long time, is evaluated as 
\begin{eqnarray}
&&\left\langle\dot{Q}_{\rm hk}\right\rangle_{\bf u}=\left\langle\overline{{W}}_{\bf u}\right\rangle_{\bf u}\nonumber\\
&=&-\displaystyle\sum_{i<j}\int d{\bf x}_{\backslash i, j}\,r_{ij}({\bf x}_{\backslash i,j}) P_{{\rm ss}\backslash i, j}({\bf x}_{\backslash i,j})\nonumber\\
&&\times\int dx_i dx_j\,\left[\frac{\partial u_i}{\partial x_j}-\frac{\partial u_j}{\partial x_i}\right]P_{{\rm ss}, i, j}(x_i, x_j)\nonumber\\
&=&\sum_{i<j}\int d{\bf x}_{\backslash i,j}\,r_{ij}({\bf x}_{\backslash i,j}) P_{{\rm ss}\backslash i, j}({\bf x}_{\backslash i, j})\nonumber\\
&&\times\int dx_i dx_j \left[u_i\frac{\partial}{\partial x_j}-u_j\frac{\partial}{\partial x_i}\right] P_{{\rm ss}, i, j}(x_i, x_j),\nonumber\\
\end{eqnarray}
where $r_{ij}({\bf x}_{\backslash i,j})$ is the rotation rate defined as $r_{ij}({\bf x}_{\backslash i,j})=\lim_{\tau\to\infty}\left\langle N_{ij}-N_{ji}\right\rangle_{\bf u}/\tau$.  
Here we assume that the stationary distribution $P_{{\rm ss}}^{\bf u}$ is independent of the additional force ${\bf u}$, i.e., $P_{\rm ss}^{\bf u}=P_{\rm ss}$, since we focus on the convergence toward a given distribution.  
The marginal distributions $P_{{\rm ss}, i, j}$ and $P_{{\rm ss}\backslash i, j}$ are defined as 
\begin{eqnarray}
P_{{\rm ss}, i, j}(x_i, x_j)&=&\int d{\bf x}_{\backslash i, j}\,P_{\rm ss}({\bf x}),\\
P_{{\rm ss}\backslash i, j}({\bf x}_{\backslash i, j})&=&\int dx_i dx_j\,P_{\rm ss}({\bf x}).
\end{eqnarray}
Note that the rotation rate $r_{ij}$ plays the role of a control parameter, which yields a steady probability current in the NESS.  
By substituting $S({\bf X})=\dot{Q}_{\rm hk}$ into Eq.~(\ref{u}) with the stationary distribution (\ref{Gibbs}), 
the heuristically found rotational force (\ref{rotational}) is reproduced with $\gamma_{ij}({\bf x}_{\backslash i,j})=-2\gamma r_{ij}({\bf x}_{\backslash i,j})$.
In addition, the KL divergence (\ref{KL}) for the optimized ${\bf u}$ relates to the time derivative of the housekeeping heat as 
\begin{eqnarray}
D[L_{{\bf u}^\gamma}|L_{\bf 0}]=\frac{1}{4T}\left\langle\dot{Q}_{\rm hk}\right\rangle_{{\bf u}^\gamma}.\label{KLtoQhk}
\end{eqnarray}
The additional force ${\bf u}^\gamma$ derived here emerges from a topological argument on the path realization ${\bf X}$.  
Such a topological effect in nonequilibrium thermodynamics was discussed by Sagawa and Hayakawa \cite{Sagawa}.  
It is pointed out that the excess entropy production generally depends on the path realization ${\bf X}$ because it is expressed in terms of a vector potential.  
However, the mathematical framework given by Sagawa and Hayakawa is applicable to arbitrary quantities dependent on a path realization, not only for the excess entropy production.  
The housekeeping heat in our case can be regarded as one of such examples, since it depends on the path realization of the closed trajectory $C$.  

\section{Exponential force}
By choosing the exponential force (\ref{exponential}) as the additional force, the dynamics (\ref{DBC_violating_Langevin}) should be solved under a periodic boundary condition.
Under a periodic boundary condition, in addition to the loop trajectory created by the rotational current, winding on the manifold of a state space can generate a closed trajectory.
Here we focus on the work performed by the nonconservative force ${\bf u}$ corresponding to the path of such a nontrivial homotopy.
For a winding trajectory, the long-time averaged work performed by the force $-{\rm grad} U+{\bf u}$ is given as 
\begin{eqnarray}
\overline{{W}}_{\bf u}=\displaystyle\lim_{\tau\to\infty}\sum_{i} \frac{N_i}{\tau} \oint_{C_i}  u_i \left({\bf x}\right) dx_i,
\end{eqnarray}
where $N_i$ denotes the winding number of the trajectory in the $i$th direction.  
The integral is taken over the single loop $C_i$ in the $i$th direction.
Note that $N_i$ is independent of $x_i$ but it may depend on the location of a winding ${\bf x}_{\backslash i}\equiv (x_1, \cdots, x_{i-1}, x_{i+1}, \cdots, x_N)^{\rm T}$.
Then the expectation of the time derivative of housekeeping heat is given as 
\begin{eqnarray}
\left\langle\dot{Q}_{\rm hk}\right\rangle_{\bf u}&=& \left\langle\overline{{W}}_{\bf u}\right\rangle_{\bf u}\nonumber\\
&=&\displaystyle\sum_i \int\, r_i({\bf x}_{\backslash i}) u_i({\bf x}) P_{{\rm ss}\backslash i}({\bf x}_{\backslash i})d{\bf x},
\end{eqnarray}
where $r_i({\bf x}_{\backslash i})$ is the winding rate in the $i$th direction defined as $r_i({\bf x}_{\backslash i})=\lim_{\tau\to\infty}\left\langle N_i\right\rangle_{\bf u}/\tau$, and $P_{{\rm ss}\backslash i}$ is the marginal distribution defined as 
\begin{eqnarray}
P_{{\rm ss}\backslash i}({\bf x}_{\backslash i})\equiv \oint_{C_i} P_{\rm ss}({\bf x}) dx_i.  
\end{eqnarray}
Similarly to the case of the rotational force, by substituting $S({\bf X})=\dot{Q}_{\rm hk}$ into Eq.~(\ref{u}), we find 
\begin{eqnarray}
u_i^\gamma({\bf x})=2T\gamma r_i({\bf x}_{\backslash i})\frac{P_{{\rm ss}\backslash i}({\bf x}_{\backslash i})}{P_{\rm ss}({\bf x})}.
\end{eqnarray}
Therefore 
the exponential force (\ref{exponential}) with $\gamma_i=2T\gamma r_i P_{{\rm ss}\backslash i}$ is reproduced from the Nemoto-Sasa theory.   
In addition, the above obtained exponential force satisfies the relation (\ref{KLtoQhk}), the same as the case of the rotational force.  
We emphasize here that our proposed exponential force is related to a nontrivial homotopy beyond that described in Ref.~\cite{Sagawa}.  
The argument on an entropy production in Ref.~\cite{Sagawa} neglected the effect of boundaries.  

\section{FDT and biased sampling}
We have found that the additional forces accelerating relaxation are derived in the framework of the variational principle.  
The cumulant generating function of the total heat in the original system, which has the trivial zero first cumulant, is given by the expectation of the housekeeping heat in the biased system.  Furthermore, Eq.~(\ref{large_deviation}) together with Eq.~(\ref{KLtoQhk}) implies that the large deviation function of the total heat in the original system is given by the expectation of the housekeeping heat in the biased system: 
\begin{eqnarray}
I_0(\overline{q})=\frac{1}{4T}\overline{q}\label{FDT}
\end{eqnarray}
with 
\begin{eqnarray}
\overline{q}=\left\langle \dot{Q}_{\rm hk}\right\rangle_{{\bf u}^\gamma}.\label{response}
\end{eqnarray}
Note that, since $\left\langle\dot{Q}_{\rm hk}\right\rangle_{{\bf u}^\gamma}\ge 0$, the large deviation function $I_0(\overline{q})$ in Eq.~(\ref{FDT}) only for non-negative $\overline{q}$ can be obtained by measuring the housekeeping heat in the biased system.  
The left-hand side of Eq.~(\ref{FDT}) denotes the intrinsic fluctuation of the time derivative of the total heat, which equals the excess heat, in the original system, namely an equilibrium state.  
On the other hand, the right-hand side with Eq.~(\ref{response}) is the response of the total heat, which equals the expectation of the housekeeping heat, in the vicinity of the additional nonconservative force ${\bf u}^\gamma$, the biased system.  
Thus Eq.~(\ref{FDT}) can be regarded as the exact form of the fluctuation-response relation in the NESS for the case that the stationary distribution is shared with equilibrium system.  
Furthermore, since $\overline{q}$ represents the energy dissipation under the perturbation ${\bf u}^\gamma$, Eq.~(\ref{FDT}) is the exact extension of the FDT in the NESS for the case that the stationary distribution is fixed.  
This is the main result of the present paper.  
We here emphasize that the derivation of Eq.~(\ref{FDT}) does not resort to a linear approximation, which usually appears in extensions of the FDT and fluctuation-response relation in the NESS.  

We finally remark the role of the additional force in the context of the so-called biased samplings before addressing the conclusion.  
Let us compare the path probabilities with and without the additional force ${\bf u}^\gamma$ as 
\begin{eqnarray}
&&\psi_{{\bf u}^\gamma}({\bf X}|{\bf x}_0)\equiv\ln\frac{L_{{\bf u}^\gamma}({\bf X}|{\bf x}_0)}{L_{\bf 0}({\bf X}|{\bf x}_0)}\nonumber\\
&&=\int_0^\tau dt\left[\frac{{\bf u}^\gamma\cdot\left(\dot{\bf x}+{\rm grad} U\right)}{2T}-\frac{\left({\bf u}^\gamma\right)^2}{4T} -\frac{{\rm div}\, {\bf u}^\gamma}{2}\right].
\end{eqnarray}
For both choices of the rotational and exponential forces, we find $\left\langle\psi_{{\bf u}^\gamma}\right\rangle_{{\bf u}^\gamma} \ge 0$ and $\left\langle\psi_{{\bf u}^\gamma}\right\rangle_{\bf 0} \le 0$.  
The first inequality represents the non-negativity of the KL divergence (\ref{KLtoQhk}) and thus holds even if ${\bf u}^\gamma$ does not coincide with our choices.  
On the other hand, the nonpositivity in the second inequality is in debt to ${\bf u}^\gamma$ of our proposal.   
Since the expectation $\langle\psi_{{\bf u}^\gamma}\rangle_{{\bf u}^\gamma}$ is realized by typical paths in the biased system with ${\bf u}^\gamma$, these two inequalities with the definition of $\psi_{{\bf u}^\gamma}$, i.e., $L_{{\bf u}^\gamma}=\exp(\psi_{{\bf u}^\gamma})L_{\bf 0}$ indicate that switches between the typical and rare path realizations are induced by the addition of our forces.  
Thus, if the typical path in the system with the DBC is the bottleneck of relaxation, such as trap at local minimum, the significant reduction of relaxation time is expected by our forces \cite{OhzekiIchiki,OhzekiIchiki2}.  
The switches between the typical and rare path realizations, which governs the relaxation, enable us to connect the fluctuation as a rare event represented by the tail of a large deviation function in the original system with the response expressed as a typical event in the biased system.  

\section{Conclusion}
We have derived the full order expression of the extension of the FDT in the NESS under a perturbation that leaves the stationary distribution unchanged. The obtained FDT has focused on the intrinsic fluctuations in an equilibrium state, while the conventional extensions of the FDT in the NESS have been focused on those in the NESS.  
The intrinsic fluctuation referred in our FDT is expressed in terms of the large deviation of the total heat in an equilibrium state.  
In addition, the conventional extensions of the FDT in the NESS refers to the ``violation of the FDT," in which they regard the response as the energy dissipation via a linear response theory and the housekeeping heat as the violating term of the FDT.  
On the other hand, our FDT assumes that the response to a nonconservative force itself is the housekeeping heat, i.e., the energy dissipation.  
From this viewpoint, the relation between the response and energy dissipation is straightforwardly given.  
Therefore only the fluctuation-response relation (\ref{FDT}) is required to obtain the extension of the FDT in our case.  
Our framework may imply that physical quantities in the NESS should be observed by measuring the difference from an equilibrium state.  
This viewpoint will allow us to give deeper understandings of thermodynamics and statistical physics in non-equilibrium states.   

\begin{acknowledgments}
M.O. acknowledges the financial support by MEXT in Japan, Grant-in Aid for Young Scientists (B) No. 24740263, Grant-in-Aid for Scientific Research (B) KAKENHI No. 15H03699, and the Kayamori Foundation of Informational Science Advancement.
\end{acknowledgments}

\end{document}